\documentclass[fleqn]{jleo}

\usepackage{amssymb,graphicx,amsfonts,bm,graphicx}
\usepackage{times,amsmath,tnqupgreek}

\usepackage{threeparttable}
\usepackage{multirow}
\usepackage{array} 
\usepackage{longtable}

\usepackage{booktabs} 

\usepackage{hyperref}
\begin{document}

\title{Digital finance, Bargaining Power and Gender Wage Gap}

\author{Qing Guo, Siyu Chen, Xiangquan Zeng\setcounter{footnote}{1}\thanks{$\langle${qingguo@ruc.edu.cn}$\rangle$
 % the footnote testing the footnote testing  the footnote testing the footnote testing\newline
\newline \hspace*{10.5pt}The author is grateful to Dr Siyu Chen and Prof. Xiangquan Zeng for useful discussions. She also
thanks Prof. Jin Feng, Prof. Xiaoning Long, Prof. Hong Liu, Prof. Fei Wang, Prof. Xiangbo Liu for helpful advice and comments. We are grateful for the advice from the 17th China Women Economists (CWE) International Workshop of 2020. 
}\\
Renmin University of China,\\ Beijing, China}
%,\\ 180 Queen's Gate,\\ London  SW7 2AZ}

\maketitle
%{\def\thefootnote{}\def\thefnmark{}
%\footnotetext{without footnote mark testing the footnote testing the footnote %testing the footnote testing the footnote testing the footnote testing the %footnote testing the footnote testing the footnote testing the footnote}}

\markboth{}{Exact Solutions To The Unsteady\\ Two-Phase Hele-Shaw Problem}

\pagestyle{headings}
\begin{abstract} 
{\textbf{Abstract:} The proliferation of internet technology has catalyzed the rapid development of digital finance, significantly impacting the optimization of resource allocation in China and exerting a substantial and enduring influence on the structure of employment and income distribution. This research utilizes data sourced from the Chinese General Social Survey and the Digital Financial Inclusion Index to scrutinize the influence of digital finance on the gender wage disparity in China. The findings reveal that digital finance reduces the gender wage gap, and this conclusion remains robust after addressing endogeneity problem using instrumental variable methods.
Further analysis of the underlying mechanisms indicates that digital finance facilitates female entrepreneurship by lowering financing barriers, thereby promoting employment opportunities for women and also empowering them to negotiate higher wages. Specially, digital finance enhances women's bargaining power within domestic settings, therefore exerts a positive influence on the wages of women. 
Sub-sample regressions demonstrate that women from economically disadvantaged backgrounds, with lower human capital, benefit more from digital finance, underscoring its inclusive nature. This study provides policy evidence for empowering vulnerable groups to increase their wages and addressing the persistent issue of gender income disparity in the labor market.
}
\end{abstract}

% The proliferation of information and communication technology has catalyzed the swift advancement of digital finance, which in turn has significant implications for wage distribution in the Chinese economy. This research utilizes data sourced from the Chinese General Social Survey and the Digital Financial Inclusion Index to scrutinize the influence of digital finance on the gender wage disparity in China.
% Our analysis reveals three key outcomes. Firstly, the advent of digital finance has contributed to the reduction of the gender wage gap. Secondly, by mitigating capital constraints and lowering operational costs, digital finance has fostered female entrepreneurship. This has not only increased employment opportunities for women but has also empowered them to negotiate higher wages and enhance their bargaining power within domestic settings. Thirdly, digital finance exerts a positive influence on the wages of women, and to a lesser extent men, from households with lower economic status. Moreover, it serves as a buffer for women against the potential wage decrement associated with childbearing and childcare responsibilities, especially under the two-child policy.
% These findings underscore the critical role digital finance plays in shaping gender dynamics within the labor market and hold substantial policy implications. They also provide empirical evidence of the amelioration of women's economic conditions, which is instrumental in bridging the gender wage gap.

\section{Introduction}

Equitable wage distribution is essential for the establishment of a labor market that is both efficient and inclusive. It constitutes a pivotal element in fostering sustainable economic expansion and ensuring societal equilibrium. Within the realm of wage distribution dynamics, the examination of the gender wage gap has garnered considerable attention from academic circles. This issue is not merely a matter of social justice but also a critical economic concern, as it directly influences the productivity and resilience of the labor force. Addressing the gender wage gap is, therefore, imperative for achieving a balanced and thriving economy. Golley et al. (2019) reveal an alarmingly high relative share of inequality of opportunity in nationwide individual labor earnings of 25 percent\cite{golley2019inequality}, and the inequality of opportunity is higher among Chinese women than among Chinese men. Deeply influenced by history and culture, Asian countries such as China face particularly conspicuous gender inequalities \cite{lee2011measures}, leading to prejudice and inefficiencies in the labor market. The exiting study suggests a negative relationship between the number of children and female labor supply\cite{ngo2020effects}. The two-child policy may further exacerbated the gender wage gap. China attaches great importance to the issues of gender equality and women’s development. A series of policies and measures has been implemented to improve gender equality, promote social justice, and improve the labor market. The global gross domestic product (GDP) growth was projected to reach US\$12 trillion by 2025 if equal gender opportunities were achieved\cite{woetzel2015power}. Thus, women’s development, and narrowing the gender gap in China  the country with one-fifth of the world’s female population  could further promote domestic and global economic development.

The evolution of information and communication technology, big data, the Internet of Things, and cloud computing has ushered in a new era of business paradigms, significantly boosting the operational efficiency within the commercial sector\cite{xu2024research,chen2021pareto, xiao2022decoupled, chen2022ba}.
% The world benefits from rapidly advancing information and communication technology, big data, the Internet of Things, and cloud computing. 
With these technological tools, the innovation of conventional finance and the development of digital finance – such as Alipay and WeChat Pay in China – have reached a new peak. With the aid of digital finance, major e-commerce platforms reached their sales peak at the 2019 Double Eleven Event, during which (based on Alipay data) over 8 million products were ordered interest-free in stages on Ant Credit Pay. Data indicates that digital finance results in significant convenience both for consumers and businesses. The inclusive nature of digital finance has attracted many scholars. Its development and popularization promote the use of financial services and reduce financial constraints, which encourages entrepreneurship among disadvantaged groups with financing difficulties small and micro businesses, and rural residents. However, the study shows that women's ventures are smaller, ‘cheaper’ to finance, and are facing difficulties of financing \cite{heilbrunn2004impact}. The study claims that the contribution to welfare resulting from female entrepreneurship to be higher than that resulting from the activity of men. However, the number of women entrepreneurs is significantly lower than that of men\cite{minniti2010female}. Clearly, women are highly entrepreneurial, but they remain at a disadvantage with regard to financing and establishing a business. Can examining whether digital finance encourages female entrepreneurship and increases women’s labor participation provide clues / explanations for the long-standing problem of the gender wage gap? This study uses 2012, 2013, and 2015 data from Chinese General Social Survey (CGSS) and the index of digital financial inclusion to ascertain how digital finance narrows the gender wage gap. It provides new empirical evidence on digital finance’s impact on the labor market in the context of promoting social equity and achieving sustainable development.

This study’s contributions are as follows. First, it provides evidence that digital finance could narrow the gender wage gap. It focuses on the impact of digital finance on women’s wages and on the alleviation of the gender wage gap – which is important in the background of digital economy and under the two-child policy. Second, it identifies that digital finance can reduce the financing threshold and the cost of entrepreneurship. This enables disadvantaged groups  especially women  to pursue entrepreneurship and drives more women into employment, thereby increasing women’s labor participation, wages, and bargaining power within the household, ultimately reducing the gender wage gap. Third, it provides evidence that digital finance could help women (but not men) with two or more pre-school children to counter the risk of a decline in wages caused by childbearing and caring under the two-child policy.

The rest of this paper is organized as follows. Section I discusses the relevant literature and logic analysis. Section II presents the dataset, variables, and the empirical strategies. Section III analyzes the empirical results. Finally, Section IV summarizes the study and presents some policy recommendations.

% \end{itemize}

\section{Literature review and conceptual analysis}

\subsection{Literature review}

The related literature mainly covers the impact of digital finance on the labor market and the gender wage gap. In China, digital finance-related research predominantly focuses on digital finance’s impact on entrepreneurship, residents’ earnings, consumption, and so on. At the micro level, digital finance promotes enterprise and individual entrepreneurship and plays an important role in the development of small and micro businesses\cite{chen2023fintech}. Through e-commerce platforms, digital finance provides small and micro businesses with access to loans, and applies risk-control technology concerning big data. This can significantly increase sales and improve commodity diversity \cite{hau2017techfin}. Moreover, the use of digital finance improves rural residents’ entrepreneurial behavior, which increases their wages and reduces the wage gap between urban and rural residents and also contributes to urbanization\cite{zhang2020trickle}. Digital finance can promote individual consumption by easing liquidity constraints and facilitating residents’ payments, especially those in rural areas, the central and western regions, and for low-income families\cite{gazel2021entrepreneurial}.

A few foreign studies have focused on the impact of digital finance on the labor market, such as digital finance promoting homogenized financial services for women, reducing financing costs, and enhancing entrepreneurship in light of limited formal financial services for women and other disadvantaged groups\cite{orser2006women}. These studies mostly examine job generation in the platform economy and the role of digital technology in replacing the labor force. It was found that the platform economy could create numerous jobs \cite{de2017impact}, promote flexible jobs (also called crowded work and the gig economy), and enrich work content, thereby increasing earning sources \cite{cantarella2021workers}. Regarding technological innovation, the digital economy has an important impact on global enterprises, skills, and employment restructuring \cite{spiezia2017jobs}. These are specifically evident with the Internet of Things, which promotes the employment of high-skilled labor and reduces the employment of low-skilled labor \cite{reimsbach2012ict,balsmeier2019time}.

The factors influencing the gender wage gap have been well researched. From the late 1980s to the mid-1990s, the gender wage gap among urban workers in China has been expanding\cite{gustafsson2000economic,maurer2002effects}, and it has continued to expand in the first decade of the 21st century \cite{demurger2007evolution}. Scholars have analyzed the causes of this gap – those that can be explained and those that cannot. The factors that can be explained include the different availability of labor, occupational gender segregation, and others. Women invest less in education, skills, and other aspects than men, and different human capital investment produces different benefits between genders\cite{becjer1964human}. There are two perspectives concerning occupational gender segregation. One is the “crowding” effect. When the proportion of women in an occupation increases, employers assume the occupation value is declining, and reduce the earning levels of this occupation\cite{bergman1974occupational}. The other is that employers prefer men when allocating labor; thus, men are more likely to get high-earning jobs\cite{reskin1990job}. Meanwhile, the factor that cannot be explained concerns gender discrimination, which is considered to be the main factor affecting the gender wage gap\cite{petersen1995separate}. Within the demographic of individuals utilizing Information and Communication Technology (ICT), the variance in wages between males and females is predominantly accounted for by the disparity in compensation for equivalent attributes\cite{moreno2008new}.

To summarize, most studies mainly focus on how digital finance promotes entrepreneurship and increases residents’ wages, while none explores whether it narrows the gender wage gap. Abundant literature exists on the factors affecting the gender wage gap  human capital, discrimination, and others. A digital economy delivers technological tools and platforms to financial innovation. The rapid development of digital finance  such as Alipay and WeChat Pay in China  is the application and implementation of information and communication technology in the field of finance. Digital finance could create a significant impact on the labor market. However, the literature on digital finance’s effect on the gender wage gap is still scarce, requiring us to examine this topic.

\subsection{Conceptual analysis}

This study expands the logic of entrepreneurial choice under liquidity constraints \cite{evans1989estimated, nguimkeu2014structural} by considering the role and influence of digital finance. At the first stage, which is before the development of digital finance, individuals had two career choices: remain a wage worker or become an entrepreneur. Becoming an entrepreneur depends on an individual’s financial constraints and entrepreneurial ability. Before digital finance had developed, the amount of capital available to entrepreneurs was less than or equal to the borrowings from their own financing channels and banks, and the maximum loan amount did not exceed the fixed multiple of the entrepreneur’s beginning-of-period wealth\cite{evans1989estimated}. Research indicates that there are barriers faced by women to access and use financial services than their male counterparts\cite{trivelli2018financial}; thus, women often face financial exclusion. Even women with certain entrepreneurial abilities were more likely to be wage workers than entrepreneurs due to financial constraints.

At the second stage, the advancement and application of information technology have spurred the robust growth of the financial and commercial sectors, giving rise to digital finance and other innovative tools\cite{chen2021improving, xiao2024simple,jiang2022role, xiao2022representation}.
In this stage, digital finance development can increase the financing channels of individuals and small and micro enterprises, reduce the financing threshold, facilitate the means of payment, and reduce operating costs\cite{hau2017techfin}. The development of digital finance can reduce the critical value of entrepreneurial ability, which helps individuals, especially women facing capital constraints, to start their own businesses. However, it has no significant impact on the entrepreneurial behavior of men who have broader financing channels and are less constrained by capital. More jobs will be created after women start businesses and promote women’s employment, thereby increasing women’s wages, leading to a reduction in the gender wage gap.

Before the development of digital finance, from a family perspective, women were more involved in home-making duties than men because of the division of labor and joint economic decisions within a household\cite{fingleton2013effects}. Women’s lower labor participation could reduce their wage levels, which could have a negative impact on their bargaining power and decision-making rights in the household. This results in an unbalanced situation with regard to wages and bargaining power between women and their husbands. Hence, when women start their businesses, it is less likely that their husbands will provide them with financial aid. They thus face financial constraints and are less likely to be entrepreneurs.

The evolution of technologies, such as big data and artificial intelligence, has led to the swift development of real economy sectors, including finance, and has increased the convenience of financing\cite{xiao2021learning, chen2024learning, chen2019deep}. The development of digital finance helps individuals, especially women, who are then significantly more likely to be financially included, which helps them decide to start their own businesses and gain employment. Hence, women’s wages are likely to increase and the wage gap between women and their husbands would reduce, especially for those women who were not previously employed or had lower wages. By increasing women’s labor participation, digital finance increases their wages and bargaining power within the household and promotes women’s development.

Based on the logic described above, we propose the following hypotheses:

\textbf{Hypothesis 1:} Digital finance can reduce the financing threshold and entrepreneurial costs such that disadvantaged groups, especially women who previously faced capital constraints, can promote women’s entrepreneurship and drive more women into employment.

\textbf{Hypothesis 2:} Digital finance has a greater effect on the improvement of women’s labor participation and wage levels compared to men, thus reducing the gender wage gap.

\textbf{Hypothesis 3: }Digital finance increases women’s wages and reduces the wage gap between women and their husbands, which can increase their bargaining power within the household.

\section{Data, variables and empirical strategies}

\subsection{Data and variable description}

Our data consist of: (i) index of digital financial inclusion constructed by the Institute of Digital finance of Peking University  used to describe the digital finance development in China; (ii) the 2012, 2013, and 2015 data published by CGSS  used to describe individual wages and individual characteristics. Among other data, the distributions of individuals’ wages and age in the CGSS data are the closest to the overall population structure of China, and the CGSS data has more information on individuals and their spouses (Luo and Li, 2019); and (iii) macro-level variables (e.g. provincial GDP per capita, provincial industrial structure, number of households using the Internet, number of households using mobile phones, urbanization rate, and development of conventional finance) from the China Statistical Yearbook and China Regional Financial Operation Report.

The research sample is the working-age population, aged 16–64 years from the CGSS database. After excluding respondents engaged mainly in farming, the total number of samples is 16,844, which includes the unemployed, self-employed, and employed. The labor wages of the unemployed are the missing value in the database, so the number of yearly wage samples is 10,803. Table 1 shows the variables’ definition and descriptive statistics.

The variables in this study are as follows:

(i) The dependent variables are ln(yearly wages) and ln(hourly wages), with wages data from the CGSS database. We take the annual labor wages (yuan) as the explained variable in the benchmark regression and use the hourly wages (yuan) for the robustness test. The hourly wages variable is the average monthly wages divided by the number of working hours per month, considering the wage difference caused by different working hours. We winsorize the annual wages and hourly wages and take the logarithm.

(ii) The core independent variable is the province-level index of digital financial inclusion. The index of digital financial inclusion is based on user data from Ant Financial and the first dimension of the index consists of three parts: extensive use of digital finance, intensive use of digital finance, and digital service provision of digital finance. The second dimension of the extensive use of digital finance is Alipay account coverage, and the intensive use of digital finance’s second dimension includes payment, lending, insurance, and investment. Digital service provision of digital finance consists of financial convenience and the cost of financial service. The general index of digital financial inclusion is obtained through the dimensionless method and the analytical hierarchy process. The dimensionless method compares the sub-indexes horizontally and vertically. 
Considering that digital finance has the characteristics of rapid expansion, the function $d=\frac{\log x-\log x^l}{\log x^h-\log x^l} \times 100$ is used to calculate and acquire the sub-indexes to mitigate the impact of extreme values and maintain the stability of the index. In the above function, $x$ is the original value of digital finance, and $x^l$ and $x^h$ represent the lower limit and the upper limit of the original value of digital finance, respectively. For a discussion of the selection of the threshold of digital finance in different conditions, see \cite{guo2020measuring}. The analytical hierarchy process is used to obtain the weight of different sub-indexes. The sub-indexes of different dimensions are calculated from the bottom to the top layer, and the general index of digital financial inclusion is synthesized using the sub-indexes and the arithmetic weighted average model. The index of digital financial inclusion of each province rose from an average of 40.004 in 2011 to 179.749 in 2014; the increase is clearly considerable. Figure 1 shows the index of digital financial inclusion of each province from 2011 to 2014.

(iii) The other explanatory variables are as follows. Based on the CGSS questionnaire and the practice used in the existing literature, this study controls for variables that may affect individual earnings: gender, age, education, marital status, health condition, nationality, Internet use, hukou, party membership, employment, and other individual characteristic variables, as well as the number of family members, number of children, and the number of the elderly in the whole family, and family economic status. Regarding employment types, consistent with previous research, we regard “sole trader (or partner),” “individual business,” and “being freelance” as entrepreneurship and the other options as employment. “Sole trader (or partner)” is defined as entrepreneurship where the individual is boss, and “individual business” and “being freelance” are defined as self-employed entrepreneurship. The study controls for provincial characteristics. Provincial GDP per capita is calculated from the provincial GDP and the provincial population, and the provincial proportion of employment in secondary and tertiary industries is used to measure the industrial structure. The provincial urbanization rate is measured by the proportion of the urban population to the total population in a province. The provincial proportion of financial institutions among country is used to measure the provincial development of conventional finance. The number of households using mobile phones is used to measure provincial mobile phone use. Provincial Internet use is measured by the number of households using the Internet. Meanwhile, we control for the year effects in all regressions.

\begin{figure}
  \centering
\caption{Index of digital financial inclusion of each province (2011–2014).}
    \includegraphics[width=1\linewidth]{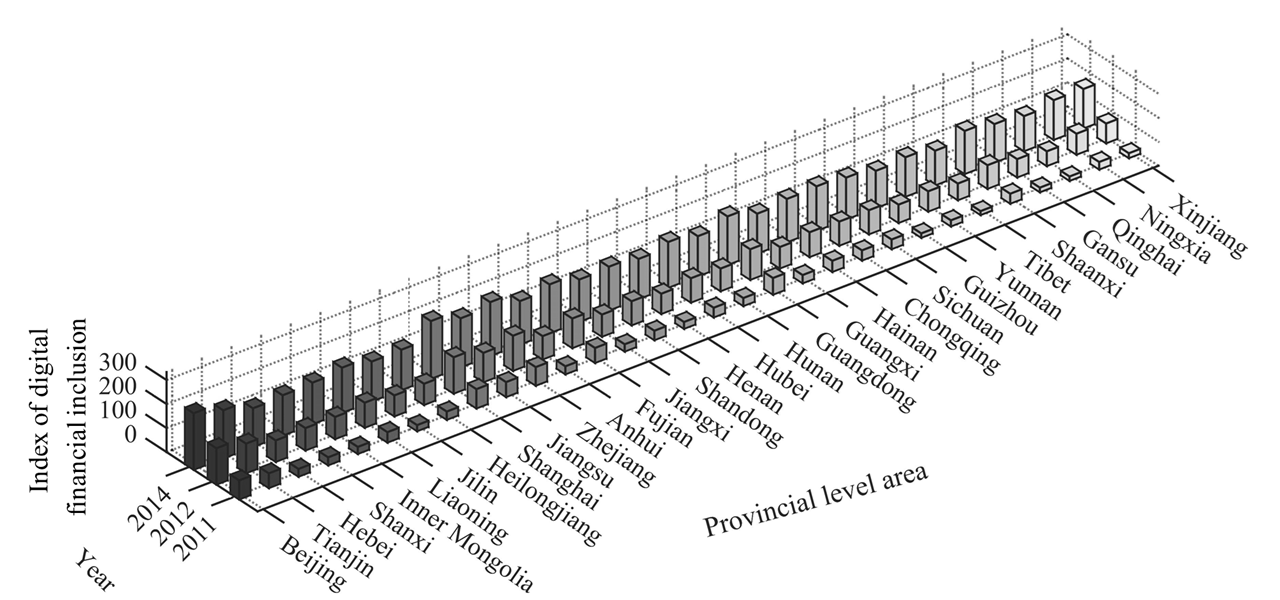} 
\end{figure}

\subsection{Empirical strategies}

\subsubsection{Benchmark regression: The impact of digital finance on gender wages}

We use the Heckman selection model to correct selection bias and estimate the impact of digital finance on the gender wage gap (Heckman, 1979). First, we construct a model measuring digital finance development's impact on labor participation and use a probit model for the regression, as in Equation (\ref{eq1}):
\begin{equation}
\operatorname{Pr}\left(\text { employ }_{i j t}=1\right)=\Phi\left(\alpha_0+\alpha_1 Z_{i j t}+\varepsilon_{i j t}\right),
    \label{eq1}
\end{equation}
where employ ${ }_{i j t}$ refers to the employment status of individual $i$ for time $t$ in province $j$, employ $_{i j t}=1$ implies employment, and employ $y_{i j t}=0$ indicates unemployment. If the annual labor wages $>0$, then employ $_{i j t}=1$; otherwise, employ ${ }_{i j t}=0 . \varphi(\cdot)$ is the standard normal cumulative distribution function. $Z_{i j t}$ refers to individual and family variables affecting employment decisions, including age, education years, marital status, family size, and number of children and the elderly, etc. $\varepsilon_{i j t}$ is the disturbance term and $\alpha_0$ is the constant term. The inverse Mill's ratio can be estimated from the model in Equation (\ref{eq1}) and is added into the wage determination equation, Equation (\ref{eq1}), as a control variable:
\begin{equation}
    \label{eq2}
\begin{gathered}
\ln \left(i n c_{i j t}\right)=\beta_0+\beta_1 F T_{j, t-1}+\beta_2 F T_{j, t-1} \times \text { gender }_{i j t} \\
+\beta_3 X_{i j t}+\beta_4 X_{i j t}^{\prime}+\sigma_{i j t}
\end{gathered}
\end{equation}

The model in Equation (\ref{eq2}), inc $c_{i j t}$ represents the wages of individual $i$ for time $t$ in province $j$, and $F T_{j, t-1}$ is the digital finance development for time $t-1$ in province $j . F T_{j, t-1} \times$ gender $_{i j t}$ is the interaction of digital finance (lagged one period) and gender. $X_{i j t}$ are the control variables including individual and family characteristics. $X_{i j t}$ are control variables including provincial and year characteristics. $\sigma_{i j t}$ is the disturbance term and $\beta_0$ is the constant term. To reduce the endogeneity problems caused by reverse causality, we add the lagged digital financial development index into the model using the 2014 digital financial index province data to match the 2015 CGSS data and so on - to analyze the impact of the previous year's digital financial development level on the current year's labor market. The impact of digital finance on the gender wage gap is shown by coefficient $\beta_2$ in model (2).

\subsubsection{Mechanisms: Digital finance and female labor participation}
The gender wage gap is explained through observable endowment factors and unobservable discrimination factors (Becker, 1964; Mincer, 1974). Digital finance reduces transaction costs and increases access to financial services, thereby alleviating women's lack of human and social capital through entrepreneurial willingness; it reduces financing discrimination, and promotes many business and employment opportunities for e-commerce and online and offline integration (Carpenter and Petersen, 2002; Wiklund and Shepherd, 2005; Bruton et al., 2015) to address unemployment and increase the wages of disadvantaged groups. Considering these, this study further discusses the impact mechanism of digital financial development on women's entrepreneurship, employment, and wages, and constructs a multinomial probit model (MNP):
\begin{equation}
\label{eq3}
\begin{gathered}
\operatorname{Pr}\left(\text { employkind }_{i j t}=k \mid x i\right)=\Phi\left(\gamma_0+\gamma_1 F T_{j, t-1}+\gamma_2 F T_{j, t-1}\right. \\
\left.\times \text { gender }_{i j t}+\gamma_3 X_{i j t}+\gamma_4 X_{i j t}^{\prime}+u_{i j t}\right) .
\end{gathered}
\end{equation}

In Equation (\ref{eq3}), employkind $_{i j t}$ refers to the employment status, and $k$ has three values (0,1, and 2). $k=0$ indicates unemployment; $k=1$ indicates employment; and $k=2$ indicates self-employment.

\section{Results and discussion}

\subsection{Benchmark regression: The impact of digital finance on wages by gender}

Based on Equation (1), the Heckman selection model is used to estimate the impact of digital finance on the gender wage gap to correct selection bias. Individual and family factors are added into the first-stage Heckman selection model, including age, education level, marital status, and family size, etc. The proportion of children who live with the interviewee includes one’s own children, stepchildren and foster children. It is added into the selection model but is not included in the wage equation. Table 2 provides the regression results of digital finance (lagged one period and normalized) and the interaction of gender and digital finance development on the ln(yearly wages), in which columns (1)–(4) gradually control the individual, family, provincial, and yearly factors. The provincial dummies are added into the model in columns (3) and (4).  The provincial dummies are controlled in the above models.

As information technology progresses, the implementation of digital technologies has swiftly overhauled the financial sector, driving its growth and enhancing the ease of its services\cite{chen2023invariant, gai2019deep, wang2020global, chen2021adaptive, lv2023ideal}. We find that, in all regressions, the coefficient of the interaction between gender and digital finance development is positive and significant. After adding more control variables, the coefficient of the interaction between gender and digital finance development increases gradually. The results indicate that the development of digital finance has a greater effect on the improvement of women's wages, which pass the robustness test. For every standard deviation of improvement in the development level of digital finance, women's wages increase 3.42–4.88 percent higher than that of men. In other words, without digital finance, China's gender wage gap may be larger. 
We find that in columns (1), (2), and (3), the coefficient of digital finance development is positive and significant at the statistical level of 1 percent, indicating that digital finance improves individuals’ wage levels. Upon adding more control variables, the coefficient of digital finance development remains positive but becomes non-significant. All the regressions reveal significantly negative gender coefficients, suggesting a substantial gender wage gap, with women’s annual wages being 37.37 to 39.26 percent lower than that of men.

The results provide evidence that digital finance can significantly improve individuals’ wages  more so for women than for men  thereby reducing the gender wage gap, and that digital finance can not only improve the efficiency of the employment market but also increases the disadvantaged group’s wages and enhances fairness in employment, which is critical to reducing the gender wage gap and promoting the sustainable and healthy operation of the labor market.

Considering individual characteristics, the coefficient of education level is significantly positive in all the regressions, indicating that an increase in the years of education can significantly increase annual wages. Older people earn significantly more than younger people, which is shown in columns (1) and (2). Married people earn less than unmarried people. The annual wages of Han residents are significantly positive in columns (1)–(4). An increase in Internet use can improve the level of an individual’s wages significantly. The coefficient of household registration is significantly negative in column (3), suggesting that the annual wages of rural households and rural-urban migrants are relatively low, consistent with existing research results. With regard to family factors, as the family size increases, the annual wages of individuals increase. However, as the number of children increases, the annual wages of individuals decline. There is a significant positive correlation between family economic status and individual annual wages. With regard to provincial factors, provincial GDP per capita has a positive effect on individuals’ wages. The development of conventional finance has a significant positive impact on annual wages. The number of households using mobile phones and the Internet has a negative impact on annual wages. The development of secondary and tertiary industries has a significant positive impact on annual wages which is shown in column (3). In all the regressions, the reversed Mills ratio is significantly negative, indicating the selection bias in the models.

To provide alternative specifications indicating that digital finance development narrows the gender wage gap, we use the sub-indices of the index of digital financial inclusion as dependent variables. The regression results are shown in Table 3. Columns (1), (2), (3) add interaction items of gender and digital finance extensive use, gender, and digital finance intensive use, gender, and digital finance digital service provision, respectively. We observe that in all regressions, the coefficient of the gender interaction item is significantly positive. For every standard deviation of improvement in the above three aspects, women's wages see an increase of 5.64, 4.49, and 2.51 percent compared to men, respectively, which indicates a greater impact of extensive and intensive use of digital finance on narrowing the gender wage gap. In the sub-indices of digital finance intensive use we find that digital payment and digital lending have a significantly positive impact on women’s wages compared with that of men. This proves that digital finance increases women’s entrepreneurship and wages by reducing capital constraints and operating costs.

\subsection{Robustness and endogeneity tests}

The results of robustness and endogeneity tests on benchmark regressions are shown in columns (1)–(4) of Table 4 and the Heckman selection model is used to estimate the results. Column (1) reports the regression results in which standard errors are clustered at the city level. We use ln(hourly wages) as the dependent variable to verify the results. Column (2) shows the results using the Heckman selection model with original standard errors, and column (3) reports the regression results in which standard errors are clustered at the city level. The provincial dummies are controlled in all the regressions. The interaction of digital finance development and gender on the wages in columns (1), (2), and (3) reveals that digital finance has a greater impact on women’s wages than men’s wages, which is consistent with the benchmark regression results.

To address the endogeneity problem resulting from reverse causality, a one-period lagged digital financial index is used in the analysis. However, the empirical analysis may still face other endogeneity problems such as missing variables. We therefore use “the spherical distance to Hangzhou city” as the instrumental variable and the Heckman selection model to conduct further analysis. The Ant Financial Services group was established in and has its headquarters in Hangzhou city, where digital finance is the most developed. Cities closer to Hangzhou should thus benefit more from the digital finance spillover. Digital finance development in China demonstrates significant spatial concentration, even after controlling for a range of economic varieties (Guo et al., 2017). This indicates that the advance and popularization of digital finance should also follow the objective law of development. The spherical distance to Hangzhou city is negatively related to digital finance development. The spherical distance to Hangzhou city has no direct impact on individuals’ wages, which meets the exogenous conditions. Column (4) of Table 4 reveals that when using the instrumental variable, the impact of digital finance on females’ wages is significantly greater than on males’ wages, which is consistent with the benchmark regression results.

\subsection{Family perspective: The impact of digital finance on wages by gender}

Considering that most of the sample was married, we further discuss the impact of digital finance on gender wage gap and women’s bargaining power within a household. First, to analyze if digital finance has a greater impact on women with lower wages within a household, we retain the married sample observations where women’s wages are equal to or lower than their husbands’ wages. The dependent variable is women’s wages minus their husbands’ wages and the wage gap is divided by 100,000 to reduce the coefficient. We firstly use the Heckman selection model to estimate this impact. For the dependent variable “women’s wages minus their husbands’ wages” is not censored because of sample selection, the model is simplified to an ordinary least squares (OLS) regression. More family and spouse factors as control variables are added to the model, including the year of marriage, age of spouse, and education level of spouse. Column (1) of Table 5 provides the regression results of digital finance (lagged one period) on the wage gap between women and their husbands. After controlling for the individual, family, and provincial factors, and the provincial and year dummies, the results show that digital finance has a significantly positive impact on improvements in the wages of women who previously had a lower economic status than their husbands. This indicates that the development of digital finance has a positive effect on narrowing the wage gap between women and their husbands within the household.

We next set aside all the married observations and use another dependent variable  the proportion of women’s wages to the couples’ wages, to measure women’s bargaining power and negotiating status in the household, with reference to Zhang and Hu (2012). Based on Equations (1) and (2), the Heckman selection model is used to estimate the impact of digital finance on the wage gap between women and their husbands. The individual, spouse and family factors are added to the first-stage Heckman selection model, including individuals’ and their spouses’ age, education level, and family size. Spouses’ hukou and job form are added to the selection model but are not included in the wage equation. We control the individual, family, and provincial factors, as well as the provincial and year dummies. The reversed Mills ratio is significantly negative, indicating the selection bias in the model. Column (2) of Table 5 provides the regression results of digital finance (lagged one period) on the proportion of women’s wages to the couples’ wages. We find that the coefficient of digital finance development is positive and significant at the 10 percent level of significance, indicating that the development of digital finance has a positive effect on increasing the proportion of women’s wages to the couples’ wages. Development of digital finance can increase the bargaining power of women within a household.

% \begin{longtable}{m{0.4\textwidth}<{\centering}m{0.15\textwidth}<{\centering}m{0.15\textwidth}<{\centering}}
% \caption{Digital finance development and gender wage gap: Family perspective}\label{tab05} \\
% \toprule
% Dependent variable：Women’s bargaining power within a household	&(1)	&(2)	 \\
% \midrule
% \endfirsthead
% \multicolumn{3}{c}%
% { Table \thetable\ continued from previous page} \\
% \toprule
% Dependent variable：Women’s bargaining power within a household	&(1)	&(2)   \\
% \midrule
% \endhead
% \bottomrule
% \multicolumn{3}{r}{{Continued on next page}} \\ 
% \endfoot

% \bottomrule
% \multicolumn{3}{l}{
%  \begin{tablenotes}[Notes]
% ***,**, and * represent significance at 1\%, 5\%, and 10\%, respectively.
% \end{tablenotes}
% }
% \endlastfoot
% % Your table data here, same as before but without the \begin{tabular} and \end{tabular} commands

% Digital finance development  &   0.2282**  &   0.0762*     \\
%   &   (0.1137)  &   (0.0461)   \\
% Individual characteristics   &   Yes  &   Yes  \\
% Family characteristics   &   Yes  &   Yes  \\
% Provincial characteristics   &   Yes  &   Yes  \\
% Year  &   Yes  &   Yes  \\
% Reversed Mills ratio  &     &   0.0304***  \\
%   &     &   (0.0055)  \\
%  \midrule
%   Observations  &   7,920  &   11,255  \\

% \end{longtable}

\subsection{Mechanism analysis}

The problem of the wage gap, or the employment opportunity inequality between genders, has long concerned the academic community. The financing threshold is still the key bottleneck for most female entrepreneurs \cite{heilbrunn2004impact}. Using the Global Findex database of 2017, the study find that formal finance in China has relatively limited inclusiveness for women\cite{qi2019measurement}. While With the advancement of information technology, digital technologies have rapidly transformed the financial industry, propelling its development and increasing its convenience\cite{gai2021multi, chen2023mapo, lv2023duet}. Digital finance and informal finance improve the possibility of using financial services for women. Hence, the rapid development of digital finance can help women finance and promote entrepreneurship, improve their wages, and drive more women into employment, and enhance overall wage levels for women. The mechanisms for digital finance to narrow the gender wage gap are as follows:

(i) Digital finance → lower financing threshold and operation cost → release entrepreneurial vitality of women with entrepreneurship → women start their own businesses and improve their earning levels → narrow the gender wage gap.

(ii) Digital finance → female entrepreneurs create more jobs → increase women's labor participation → increase women's wages → a narrower gender wage gap.

% \begin{longtable}{m{0.4\textwidth}<{\centering}m{0.15\textwidth}<{\centering}m{0.15\textwidth}<{\centering}}
% \caption{Gender differences in digital finance, self-employment, and employment}\label{tab06} \\
% \toprule
% Dependent variable：Labor participation	&(1)Self-employed	&(2)Employed		 \\
% \midrule
% \endfirsthead
% \multicolumn{3}{c}%
% { Table \thetable\ continued from previous page} \\
% \toprule
% Dependent variable：Labor participation	&(1)Self-employed	&(2)Employed   \\
% \midrule
% \endhead
% \bottomrule
% \multicolumn{3}{r}{{Continued on next page}} \\ 
% \endfoot

% \bottomrule
% \multicolumn{3}{l}{
%  \begin{tablenotes}[Notes]
% ***,**, and * represent significance at 1\%, 5\%, and 10\%, respectively.
% \end{tablenotes}
% }
% \endlastfoot
% % Your table data here, same as before but without the \begin{tabular} and \end{tabular} commands

% Gender (female) × Digital finance  &   0.0825**  &   0.1087***   \\
%   &   (0.0385)  &   (0.0321)   \\
% Digital finance  &   0.8297*  &   0.5801   \\
%   &   (0.4384)  &   (0.3564)   \\
% Individual characteristics  &   Yes  &   Yes   \\
% Family characteristics  &   Yes  &   Yes   \\
% Provincial characteristics   &   Yes  &   Yes   \\
% Year   &   Yes  &   Yes   \\
%  \midrule
%  Observations  &   16,566   &   \\
% Chi-square  &   4,850.254   &   \\

% \end{longtable}

Previous studies have found that women's financial literacy levels and the proportion of women with financial accounts are lower than those of men\cite{woodyard2012financial}. This suggests that women are subject to financial constraints, and entrepreneurship is restricted for women. Technological advancements, particularly in the realms of big data and artificial intelligence, have catalyzed the rapid expansion of tangible economic industries, including the financial sector, while simultaneously enhancing the ease with which financing can be obtained\cite{hau2017techfin, chen2021deep, chen2021multi, chen2024pareto}. The development of digital finance can improve the availability of financial services, reduce the financing threshold, and promote women’s entrepreneurship. We use an MNP to estimate the probability of entrepreneurship and employment for women compared to men. The provincial dummies are controlled in all the regressions.

Table 6 reports the gender heterogeneity of the impact of digital finance development on entrepreneurship. The results show that with the control variables (i.e. individual, family, provincial, and year characteristics), digital finance promotes women’s entrepreneurship at a statistically significant level of 5 percent compared with men. This is because the financing threshold is high under traditional financing channels, and women's financing is limited, while the development of digital finance reduces the financing threshold and releases women’ s entrepreneurial vitality. By contrast, men are less constrained by financing and can obtain a wide range of financial services for entrepreneurship on their own.

In comparison to employment for men, digital finance has a significant positive impact on female employment, at a significance level of 1 percent. The results are robust when the standard errors are clustered at the city level. Female entrepreneurs are more likely to accept and hire female employees in businesses (Allen and Langowitz, 2003). Hence, digital finance promotes women’s entrepreneurship and creates more jobs to improve women’s labor participation.

% \begin{longtable}{m{0.4\textwidth}<{\centering}m{0.15\textwidth}<{\centering}m{0.15\textwidth}<{\centering}}
% \caption{ Gender differences in digital finance and different forms of employment}\label{tab06} \\
% \toprule
% Dependent variable：Forms of employment	&(1)Standard form of employment	&(2)Standard form of employment	 \\
% \midrule
% \endfirsthead
% \multicolumn{3}{c}%
% { Table \thetable\ continued from previous page} \\
% \toprule
% Dependent variable：Forms of employment	&(1)Standard form of employment	&(2)Standard form of employment   \\
% \midrule
% \endhead
% \bottomrule
% \multicolumn{3}{r}{{Continued on next page}} \\ 
% \endfoot

% \bottomrule
% \multicolumn{3}{l}{
%  \begin{tablenotes}[Notes]
% ***,**, and * represent significance at 1\%, 5\%, and 10\%, respectively.
% \end{tablenotes}
% }
% \endlastfoot
% % Your table data here, same as before but without the \begin{tabular} and \end{tabular} commands

% Gender (female) × Digital finance  &   0.1007***  &   0.1220***    \\
%   &   (0.0384)  &   (0.0356)   \\
% Digital finance  &   0.7666*  &   0.7102*  \\
%   &   (0.4184)  &   (0.4042)   \\
% Individual characteristics  &   Yes  &   Yes   \\
% Family characteristics  &   Yes  &   Yes   \\
% Provincial characteristics   &   Yes  &   Yes   \\
% Year   &   Yes  &   Yes   \\
%  \midrule
% Observations  &   14,102  &   \\
% Chi-square  &   4,721.476 & 

% \end{longtable}

To ascertain the source of employment, we further divide the employed into those in non-standard employment and those in standard employment. We firstly define standard forms of employment and then reversely deduct the scope of non-standard forms of employment. We define standard employment in China’s labor market as employment with high employment stability and social welfare security. Specifically, if the firm or employer signs a labor contract for the workers and provides them with endowment insurance, the workers are regarded as having standard employment; otherwise, they are classified as non-standard employment. The results are shown in Table 7. We find that compared to men, digital finance promotes women’s standard employment and non-standard employment at a statistically significant level of 1 percent, and the results are robust when the standard errors are clustered at the city level.

Another possible mechanism is that digital finance development generates more jobs in the platform and gig economies, as in the case of Uber drivers, live streamers, and food delivery workers, which are time-flexible and non-labor-intensive jobs. These jobs do not provide labor contracts and endowment insurance. The study finds that jobs in the platform economy represented by live streamers make it possible for women to achieve a work life balance, which could increase women’s wages and narrow the gender wage gap\cite{qi2019measurement}. However, at present CGSS data do not reveal if the individuals are working in the platform economy. Hence, if digital finance can directly promote jobs in the platform economy, additional data are required to conduct a further study. Policy documents by the National Development and Reform Commission encourage individuals engaged in platform labor to register individual businesses recently, so that, in future, self-employment could be a way to measure platform jobs.

\subsection{Heterogeneity of the impact of digital finance on gender wage gap}

To deepen understanding of the relationship between digital finance and the reduction of the gender wage gap, this study further examined the groups of women that benefit more from the development of digital finance. The existing literature shows that differences in the personal endowment and financial threshold between men and women are the predominant causes of the restrictions on women’s financing\cite{sioson2019closing}. With the development of science and technology and the popularization of digital finance, the availability of finance has improved and the threshold of financing is lowered, which can improve women’s ability to obtain financing, promote women’s entrepreneurship, enhance their employment, and increase their wages. Here, we further study the heterogeneous effect of digital finance on women’s wages from the aspects of their family economic status, human capital, and pre-school children.

The results of the heterogeneity analysis of women and men are based on the Heckman selection model and are reported in Table 8. Individual factors are added into the first-stage Heckman selection model, including age, education years, marital status, health conditions, family size, and number of children, and the elderly. Regarding family economic status, digital finance has a significant positive impact on the wages of women (and men) at a lower family economic status level, as the coefficients of the interaction between digital finance development and family economic status (low) in columns (1) and (2) of Table 8 are both significantly positive. The results can be explained as follows. digital finance can broaden the financing channels of women (and men) who have a lower family economic status level, provide more extensive and convenient financing services and information, and promote women (and men)’s inclusion to enhance the wages of women with low economic status compared with those with higher economic status. The seemingly unrelated estimation (SUEST) shows that the coefficients among the group of women and men with lower economic status have significant differences (F = 113.06, p = 0.0870). With digital finance development, the gender wage gap can be narrowed among the group with a lower family economic status.

Columns (3) and (4) of Table 8 reveal that digital finance benefits women with lower human capital, as the coefficient of the interaction between digital finance development and human capital (low) is significantly positive among women. This is because digital finance includes disadvantaged people, especially women with low education levels. As digital finance is easy to operate, it is easier to include women with low education levels who had financial constraints before. In comparison, men are more likely to be financially included, so digital finance does not show a significant difference among men with lower education and higher education.

Meanwhile, there are some interesting results in terms of pre-school children. Column (5) of Table 8 shows that digital finance has a significant positive impact on women with two or more pre-school children (compared with women with no pre-school children or one pre-school child), as the coefficient of the interaction between digital finance development and pre-school children (two or more) is significantly positive. It indicates that digital finance can help women with two or more pre-school children (compared with women with no pre-school children or one pre-school child) increase their wages, countering the risk of a decline in wages caused by women’s childbearing and caring after the implementation of the two-child policy. This policy has placed a greater burden on female workers and resulted in negative effects on their wages compared to men (Yang and Zhou, 2019). However, digital finance shows no significant impact on men with two or more pre-school children, as seen in column (6), as men assume less responsibility with respect to caring and childbearing.

In summary, the heterogeneity analysis reveals that digital finance can help women with low education levels, those with a low family economic status, and those with two or more pre-school children enjoy greater benefits, which reflects the inclusive effects of digital finance. By comparison, digital finance can also help men with a lower family economic status but does not show a significant impact on men with a low education level and those with two or more pre-schoolers. Hence, digital finance can narrow the gender wage gap in the group with lower human capital and lower family economic status and counter the risk of decline in women’ wages caused by to their childbearing and caring after the implementation of the two-child policy.

\section{Conclusions and implications}

Based on the 2012, 2013, and 2015 data from the CGSS and the index of digital financial inclusion, this study examined the impact of digital finance development on the gender wage gap. The results show that, first, women’s wages recorded an increase of 3.42–4.88 percent compared to those of men for every standard deviation of digital finance growth, which can narrow the gender wage gap. Second, through digital finance, the financing threshold and the cost of entrepreneurship are reduced and women’s entrepreneurship and employment are promoted, thereby improving their labor participation and wages, which leads to a reduction in the gender wage gap and stronger bargaining power for women within the household. Third, digital finance has a significant impact on the wages of women (and men) who have a lower family economic status and on women with lower human capital. It also helps women (but not men) counter the risk of a decline in wages due to childbearing and caring under the two-child policy, which reflects the inclusion effects of digital finance. The study’s limitation lies in that the CGSS data do not reveal if the individuals are working in the platform economy. Hence, if digital finance can promote platform gigs or jobs directly and reduce the gender wage gap, additional data are required to conduct a further analysis. This study highlights the importance of digital finance in easing the gender wage gap in China’s labor market, alleviating the impact of the two-child policy on women’s employment, promoting the employment of disadvantaged groups, and improving employment conditions in China. In the context of the digital economy, to promote gender equality in the labor market and narrow the gender wage gap, we should continue to strengthen network construction, expand the coverage of digital finance, and promote digital finance’s inclusive effects to encourage women’s innovation and entrepreneurship vitality. Simultaneously, we should expand education and training, improve individuals’ financial literacy, and promote digital finance’s availability.

\bibliographystyle{unsrt}

\bibliography{BibFile}

\begin{thebibliography}{10}

\bibitem{golley2019inequality}
Jane Golley, Yixiao Zhou, and Meiyan Wang.
\newblock Inequality of opportunity in china's labor earnings: The gender dimension.
\newblock {\em China \& World Economy}, 27(1):28--50, 2019.

\bibitem{lee2011measures}
Jae~Kyung Lee and Hye-Gyong Park.
\newblock Measures of women's status and gender inequality in asia: issues and challenges.
\newblock {\em Asian Journal of Women's Studies}, 17(2):7--31, 2011.

\bibitem{ngo2020effects}
Anh~P Ngo.
\newblock Effects of vietnam’s two-child policy on fertility, son preference, and female labor supply.
\newblock {\em Journal of Population Economics}, 33(3):751--794, 2020.

\bibitem{woetzel2015power}
Jonathan Woetzel.
\newblock The power of parity: How advancing women's equality can add $12 trillion to global growth.
\newblock Technical report, 2015.

\bibitem{xu2024research}
Ting Xu, Iris Li, Qishi Zhan, Yuxiang Hu, and Haowei Yang.
\newblock Research on intelligent system of multimodal deep learning in image recognition.
\newblock {\em Journal of Computing and Electronic Information Management}, 12(3):79--83, 2024.

\bibitem{chen2021pareto}
Zhengyu Chen, Jixie Ge, Heshen Zhan, Siteng Huang, and Donglin Wang.
\newblock Pareto self-supervised training for few-shot learning.
\newblock In {\em Proceedings of the IEEE/CVF Conference on Computer Vision and Pattern Recognition}, pages 13663--13672, 2021.

\bibitem{xiao2022decoupled}
Teng Xiao, Zhengyu Chen, Zhimeng Guo, Zeyang Zhuang, and Suhang Wang.
\newblock Decoupled self-supervised learning for graphs.
\newblock {\em Advances in Neural Information Processing Systems}, 35:620--634, 2022.

\bibitem{chen2022ba}
Zhengyu Chen, Teng Xiao, and Kun Kuang.
\newblock Ba-gnn: On learning bias-aware graph neural network.
\newblock In {\em 2022 IEEE 38th International Conference on Data Engineering (ICDE)}, pages 3012--3024. IEEE, 2022.

\bibitem{heilbrunn2004impact}
Sibylle Heilbrunn.
\newblock Impact of gender on difficulties faced by entrepreneurs.
\newblock {\em The International Journal of Entrepreneurship and Innovation}, 5(3):159--165, 2004.

\bibitem{minniti2010female}
Maria Minniti.
\newblock Female entrepreneurship and economic activity.
\newblock {\em The European Journal of Development Research}, 22:294--312, 2010.

\bibitem{chen2023fintech}
Siyu Chen and Qing Guo.
\newblock Fintech, strategic incentives and investment to human capital, and mses innovation.
\newblock {\em The North American Journal of Economics and Finance}, 68:101963, 2023.

\bibitem{hau2017techfin}
Harald Hau, Yi~Huang, Hongzhe Shan, and Zixia Sheng.
\newblock Techfin in china: Credit market completion and its growth effect.
\newblock In {\em ABFER 6th Annual Conference, Singapore}, 2017.

\bibitem{zhang2020trickle}
Xun Zhang, Ying Tan, Zonghui Hu, Chen Wang, and Guanghua Wan.
\newblock The trickle-down effect of fintech development: From the perspective of urbanization.
\newblock {\em China \& World Economy}, 28(1):23--40, 2020.

\bibitem{gazel2021entrepreneurial}
Marco Gazel and Armin Schwienbacher.
\newblock Entrepreneurial fintech clusters.
\newblock {\em Small Business Economics}, 57:883--903, 2021.

\bibitem{orser2006women}
Barbara~J Orser, Allan~L Riding, and Kathryn Manley.
\newblock Women entrepreneurs and financial capital.
\newblock {\em Entrepreneurship Theory and practice}, 30(5):643--665, 2006.

\bibitem{de2017impact}
Willem~Pieter De~Groen, Zachary Kilhoffer, Karolien Lenaerts, and Nicolas Salez.
\newblock The impact of the platform economy on job creation.
\newblock {\em Intereconomics}, 52:345--351, 2017.

\bibitem{cantarella2021workers}
Michele Cantarella and Chiara Strozzi.
\newblock Workers in the crowd: the labor market impact of the online platform economy.
\newblock {\em Industrial and Corporate Change}, 30(6):1429--1458, 2021.

\bibitem{spiezia2017jobs}
Vincenzo Spiezia.
\newblock Jobs and skills in the digital economy.
\newblock 2017.

\bibitem{reimsbach2012ict}
Christian Reimsbach-Kounatze and Cristina~Serra Vallejo.
\newblock {\em ICT Skills and Employment: New Competences and Jobs for a Greener and Smarter Economy}.
\newblock OECD, 2012.

\bibitem{balsmeier2019time}
Benjamin Balsmeier and Martin Woerter.
\newblock Is this time different? how digitalization influences job creation and destruction.
\newblock {\em Research policy}, 48(8):103765, 2019.

\bibitem{gustafsson2000economic}
Bj{\"o}rn Gustafsson and Shi Li.
\newblock Economic transformation and the gender earnings gap in urban china.
\newblock {\em Journal of Population Economics}, 13:305--329, 2000.

\bibitem{maurer2002effects}
Margaret Maurer-Fazio and James Hughes.
\newblock The effects of market liberalization on the relative earnings of chinese women.
\newblock {\em Journal of Comparative Economics}, 30(4):709--731, 2002.

\bibitem{demurger2007evolution}
Sylvie D{\'e}murger, Martin Fournier, and Yi~Chen.
\newblock The evolution of gender earnings gaps and discrimination in urban china, 1988--95.
\newblock {\em The Developing Economies}, 45(1):97--121, 2007.

\bibitem{becjer1964human}
GS~Becjer.
\newblock {\em Human capital: A theoretical and empirical analysis}, volume~2.
\newblock Columbia University Press, 1964.

\bibitem{bergman1974occupational}
Barbara Bergman.
\newblock Occupational segregation, wages and profits when employers discriminate by race and sex.
\newblock {\em Eastern Economic Journal}, 1(1-2):103--110, 1974.

\bibitem{reskin1990job}
Barbara Reskin and Patricia~A Roos.
\newblock {\em Job queues, gender queues: Explaining women's inroads into male occupations}.
\newblock Temple University Press, 1990.

\bibitem{petersen1995separate}
Trond Petersen and Laurie~A Morgan.
\newblock Separate and unequal: Occupation-establishment sex segregation and the gender wage gap.
\newblock {\em American Journal of Sociology}, 101(2):329--365, 1995.

\bibitem{moreno2008new}
Eva Moreno-Galbis and Fran{\c{c}}ois-Charles Wolff.
\newblock New technologies and the gender wage gap: Evidence from france.
\newblock {\em Relations industrielles}, 63(2):317--339, 2008.

\bibitem{evans1989estimated}
David~S Evans and Boyan Jovanovic.
\newblock An estimated model of entrepreneurial choice under liquidity constraints.
\newblock {\em Journal of political economy}, 97(4):808--827, 1989.

\bibitem{nguimkeu2014structural}
Pierre Nguimkeu.
\newblock A structural econometric analysis of the informal sector heterogeneity.
\newblock {\em Journal of Development Economics}, 107:175--191, 2014.

\bibitem{trivelli2018financial}
Carolina Trivelli, BID Invest, Marcela Marincioni, FINDO~Jaqueline Pels, and ENI Di~Tella.
\newblock Financial inclusi{\'o}n for women: a way forward.
\newblock {\em Gender economic equity: An imperative for the G20}, 88, 2018.

\bibitem{chen2021improving}
Zhengyu Chen, Donglin Wang, and Shiqian Yin.
\newblock Improving cold-start recommendation via multi-prior meta-learning.
\newblock In {\em Advances in Information Retrieval: 43rd European Conference on IR Research, ECIR 2021, Virtual Event, March 28--April 1, 2021, Proceedings, Part II 43}, pages 249--256. Springer, 2021.

\bibitem{xiao2024simple}
Teng Xiao, Huaisheng Zhu, Zhengyu Chen, and Suhang Wang.
\newblock Simple and asymmetric graph contrastive learning without augmentations.
\newblock {\em Advances in Neural Information Processing Systems}, 36, 2024.

\bibitem{jiang2022role}
Yinjie Jiang, Zhengyu Chen, Kun Kuang, Luotian Yuan, Xinhai Ye, Zhihua Wang, Fei Wu, and Ying Wei.
\newblock The role of deconfounding in meta-learning.
\newblock In {\em International Conference on Machine Learning}, pages 10161--10176. PMLR, 2022.

\bibitem{xiao2022representation}
Teng Xiao, Zhengyu Chen, and Suhang Wang.
\newblock Representation matters when learning from biased feedback in recommendation.
\newblock In {\em Proceedings of the 31st ACM International Conference on Information \& Knowledge Management}, pages 2220--2229, 2022.

\bibitem{fingleton2013effects}
Bernard Fingleton and Simonetta Longhi.
\newblock The effects of agglomeration on wages: Evidence from the micro-level.
\newblock {\em Journal of Regional Science}, 53(3):443--463, 2013.

\bibitem{xiao2021learning}
Teng Xiao, Zhengyu Chen, Donglin Wang, and Suhang Wang.
\newblock Learning how to propagate messages in graph neural networks.
\newblock In {\em Proceedings of the 27th ACM SIGKDD Conference on Knowledge Discovery \& Data Mining}, pages 1894--1903, 2021.

\bibitem{chen2024learning}
Zhengyu Chen, Teng Xiao, Kun Kuang, Zheqi Lv, Min Zhang, Jinluan Yang, Chengqiang Lu, Hongxia Yang, and Fei Wu.
\newblock Learning to reweight for generalizable graph neural network.
\newblock In {\em Proceedings of the AAAI Conference on Artificial Intelligence}, volume~38, pages 8320--8328, 2024.

\bibitem{chen2019deep}
Zhengyu Chen, Sibo Gai, and Donglin Wang.
\newblock Deep tensor factorization for multi-criteria recommender systems.
\newblock In {\em 2019 IEEE International Conference on Big Data (Big Data)}, pages 1046--1051. IEEE, 2019.

\bibitem{guo2020measuring}
Feng Guo, Jingyi Wang, Fang Wang, Tao Kong, Xun Zhang, and Zhiyun Cheng.
\newblock Measuring china’s digital financial inclusion: Index compilation and spatial characteristics.
\newblock {\em China Economic Quarterly}, 19(4):1401--1418, 2020.

\bibitem{chen2023invariant}
Zhengyu Chen, Yishu Gong, Liangliang Yang, Jianyu Zhang, Wei Zhang, Sihong He, and Xusheng Zhang.
\newblock Invariant graph neural network for out-of-distribution nodes.
\newblock In {\em Proceedings of the 2023 15th International Conference on Machine Learning and Computing}, pages 192--196, 2023.

\bibitem{gai2019deep}
Sibo Gai, Feng Zhao, Yachen Kang, Zhengyu Chen, Donglin Wang, and Ao~Tang.
\newblock Deep transfer collaborative filtering for recommender systems.
\newblock In {\em PRICAI 2019: Trends in Artificial Intelligence: 16th Pacific Rim International Conference on Artificial Intelligence, Cuvu, Yanuca Island, Fiji, August 26-30, 2019, Proceedings, Part III 16}, pages 515--528. Springer, 2019.

\bibitem{wang2020global}
Shuliang Wang, Jingting Yang, Zhengyu Chen, Hanning Yuan, Jing Geng, and Zhen Hai.
\newblock Global and local tensor factorization for multi-criteria recommender system.
\newblock {\em Patterns}, 1(2), 2020.

\bibitem{chen2021adaptive}
Shiqi Chen, Zhengyu Chen, and Donglin Wang.
\newblock Adaptive adversarial training for meta reinforcement learning.
\newblock In {\em 2021 International Joint Conference on Neural Networks (IJCNN)}, pages 1--8. IEEE, 2021.

\bibitem{lv2023ideal}
Zheqi Lv, Zhengyu Chen, Shengyu Zhang, Kun Kuang, Wenqiao Zhang, Mengze Li, Beng~Chin Ooi, and Fei Wu.
\newblock Ideal: Toward high-efficiency device-cloud collaborative and dynamic recommendation system.
\newblock {\em arXiv preprint arXiv:2302.07335}, 2023.

\bibitem{qi2019measurement}
H~Qi and Z~Li.
\newblock Measurement and evaluation of financial inclusion in china.
\newblock {\em J Quant Technol Econ}, 36:101--17, 2019.

\bibitem{gai2021multi}
Sibo Gai, Zhengyu Chen, and Donglin Wang.
\newblock Multi-modal meta continual learning.
\newblock In {\em 2021 International Joint Conference on Neural Networks (IJCNN)}, pages 1--8. IEEE, 2021.

\bibitem{chen2023mapo}
Yuyan Chen, Zhihao Wen, Ge~Fan, Zhengyu Chen, Wei Wu, Dayiheng Liu, Zhixu Li, Bang Liu, and Yanghua Xiao.
\newblock Mapo: Boosting large language model performance with model-adaptive prompt optimization.
\newblock In {\em Findings of the Association for Computational Linguistics: EMNLP 2023}, pages 3279--3304, 2023.

\bibitem{lv2023duet}
Zheqi Lv, Wenqiao Zhang, Shengyu Zhang, Kun Kuang, Feng Wang, Yongwei Wang, Zhengyu Chen, Tao Shen, Hongxia Yang, Beng~Chin Ooi, et~al.
\newblock Duet: A tuning-free device-cloud collaborative parameters generation framework for efficient device model generalization.
\newblock In {\em Proceedings of the ACM Web Conference 2023}, pages 3077--3085, 2023.

\bibitem{woodyard2012financial}
Ann Woodyard and Cliff Robb.
\newblock Financial knowledge and the gender gap.
\newblock {\em Journal of Financial Therapy}, 3(1):1, 2012.

\bibitem{chen2021deep}
Zhengyu Chen, Ziqing Xu, and Donglin Wang.
\newblock Deep transfer tensor decomposition with orthogonal constraint for recommender systems.
\newblock In {\em Proceedings of the AAAI Conference on Artificial Intelligence}, volume~35, pages 4010--4018, 2021.

\bibitem{chen2021multi}
Zhengyu Chen and Donglin Wang.
\newblock Multi-initialization meta-learning with domain adaptation.
\newblock In {\em ICASSP 2021-2021 IEEE International Conference on Acoustics, Speech and Signal Processing (ICASSP)}, pages 1390--1394. IEEE, 2021.

\bibitem{chen2024pareto}
Zhengyu Chen, Teng Xiao, Donglin Wang, and Min Zhang.
\newblock Pareto graph self-supervised learning.
\newblock In {\em ICASSP 2024-2024 IEEE International Conference on Acoustics, Speech and Signal Processing (ICASSP)}, pages 6630--6634. IEEE, 2024.

\bibitem{sioson2019closing}
Erica~Paula Sioson.
\newblock Closing the gender gap in financial inclusion through fintech.
\newblock 2019.

\end{thebibliography}

\end{document}